\documentclass[3p]{elsarticle}

\usepackage{lineno,hyperref}
\usepackage{gensymb}
\usepackage{amsmath}
\usepackage{booktabs,caption}
\usepackage[flushleft]{threeparttable}
\usepackage[justification=centering]{caption}
\usepackage{xcolor}
\usepackage{graphicx}
\usepackage{multicol}
\usepackage{dcolumn}
\usepackage{multirow}
\usepackage{xspace}
\usepackage{mathrsfs}
\usepackage{xfrac}
\usepackage{braket}
\usepackage{xifthen}
\usepackage{rotating}
\usepackage{longtable}
\usepackage{chemformula}
\let\ce\ch

\DeclareSymbolFont{matha}{OML}{txmi}{m}{it}
\newcommand{\wn}{cm$^{-1}$\xspace} 
\newcommand{\mrm}[1]{\ensuremath{\mathrm{#1}}} 

\renewcommand{\bra}[2][]{%
	\ifthenelse{\isempty{#1}}%
	{\ensuremath{\left\langle #2 \right\rvert}}%
	{\ensuremath{_{#1}\left\langle #2 \right\rvert}}}
\renewcommand{\ket}[2][]{%
	\ifthenelse{\isempty{#1}}%
	{\ensuremath{\left\lvert #2 \right\rangle}}%
	{\ensuremath{\left\lvert #2 \right\rangle_{#1}}}}

\newcolumntype{.}{D{.}{.}{-1}}
\newcolumntype{m}{D{.}{.}{10}}
\newcolumntype{d}[1]{D{.}{.}{#1}}

\journal{Journal of Quantitative Spectroscopy \& Radiative Transfer}

\begin{document}
	
	\begin{frontmatter}
		
		\title{Extensive ro-vibrational analysis of deuterated-cyanoacetylene (\ce{DC3N}) from millimeter-wavelengths to the infrared domain}
		\tnotetext[mytitlenote]{Supplementary material available.}

		\author[ciamician]{Mattia Melosso\corref{mycorrespondingauthor1}}
		\cortext[mycorrespondingauthor1]{Corresponding authors}
		\ead{mattia.melosso2@unibo.it}
		\author[mpe]{Luca Bizzocchi}
		\author[industriale]{Aleksandra Adamczyk}
		\author[industriale]{Elisabetta Can\`{e}}
		\author[mpe]{Paola Caselli}
		\author[arcetri]{Laura Colzi}
		\author[ciamician]{Luca Dore}
		\author[mpe]{Barbara M. Giuliano}
		\author[rennes]{Jean-Claude Guillemin}
		\author[ismo]{Marie-Aline Martin-Drumel}
		\author[ismo,soleil]{Olivier Pirali}
		\author[venezia]{Andrea Pietropolli Charmet}
		\author[mpe]{Domenico Prudenzano}
		\author[arcetri]{V\'ictor M. Rivilla}
		\author[industriale]{Filippo Tamassia\corref{mycorrespondingauthor1}}
		\ead{filippo.tamassia@unibo.it}
		
		\address[ciamician]{Dipartimento di Chimica ``Giacomo Ciamician'', Universit\`a di Bologna, Via F.~Selmi 2, 40126 Bologna (Italy)}
		\address[mpe]{Center for Astrochemical Studies, Max Planck Institut f\"ur extraterrestrische Physik Gie\ss enbachstra\ss e 1, 85748 Garching bei M\"unchen (Germany)}
		\address[industriale]{Dipartimento di Chimica Industriale ``Toso Montanari'', Universit\`a di Bologna, Viale del Risorgimento 4, 40136 Bologna (Italy)}
		\address[arcetri]{INAF-Osservatorio Astrofisico di Arcetri, Largo Enrico Fermi 5, 50125, Firenze (Italy)}
		\address[rennes]{Univ Rennes, Ecole Nationale Sup\'{e}rieure de Chimie de Rennes, CNRS, ISCR--UMR6226, 35000 Rennes (France)}
		\address[ismo]{Universit\'e Paris-Saclay, CNRS, Institut des Sciences Mol\'eculaires d'Orsay, 91405 Orsay (France)}
		\address[soleil]{SOLEIL Synchrotron, AILES beamline, l'Orme des Merisiers, Saint-Aubin, 91190 Gif-sur-Yvette, (France)}
		\address[venezia]{Dipartimento di Scienze Molecolari e Nanosistemi, Universit\`a Ca' Foscari Venezia, Via Torino 155, 30172 Mestre (Italy)}
		
		
		\begin{abstract}
			
			Cyanoacetylene, the simplest cyanopolyyne, is an abundant interstellar molecule commonly observed in a vast variety of astronomical sources. Despite its importance as a potential tracer of the evolution of star-forming processes, the deuterated form of cyanoacetylene is less observed and less studied in the laboratory than the main isotopologue.
			Here, we report the most extensive spectroscopic characterization of \ce{DC3N} to date, from the millimeter domain to the infrared region.
			Rotational and ro-vibrational spectra have been recorded using millimeter-wave frequency-modulation and Fourier-transform infrared spectrometers, respectively.
			All the vibrational states with energy up to 1015\,\wn have been analyzed in a combined fit, where the effects due to anharmonic resonances have been adequately accounted for. The analysis contains over 6500 distinct transition frequencies, from which all the vibrational energies have been determined with good precision for many fundamental, overtone, and combination states.
			This work provides a comprehensive line catalog for astronomical observations of \ce{DC3N}.
			
		\end{abstract}
		
		\begin{keyword}
			Cyanoacetylene \sep Interstellar species \sep Ro-vibrational spectroscopy \sep Spectral analysis \sep Anharmonic resonances \sep Line catalog
		\end{keyword}
		
		
	\end{frontmatter}
		
	\section{Introduction}\label{sec:intro}
	
	Highly unsaturated molecules account for a large portion of the known interstellar species \cite{mcguire2018census}.
	For instance, the presence of several carbon-chain molecules is one of the most characteristic features of the chemical composition of starless cores, such as the Taurus Molecular Cloud (TMC-1), one of the brightest source of carbon-chain species \cite{yamamoto2017book}.
	Among the unsaturated molecular species, cyanopolyynes, i.e., linear molecules of general chemical formula \ce{HC_{2n+1}N}, are widespread in the interstellar medium (ISM) and all members up to \ce{HC11N} have been detected to date \cite{Loomis2020}.
	Cyanoacetylene (\ce{HC3N}, IUPAC name prop-2-ynenitrile), the simplest member of the cyanopolyynes family, was found to be an abundant species in a large variety of astronomical objects: starless cores \cite{suzuki1992dark}, post-AGB objects \cite{wyrowski2003crl}, carbon-rich circumstellar envelopes \cite{decin2010warm}, massive star-forming regions \cite{li2011large}, protoplanetary disks \cite{chapillon2012proto}, solar-type protostars \cite{al2017iras}, external galaxies \cite{rico2020super}, and Galactic Center molecular clouds \cite{zeng2018complex}.
	
	The deuterated form of cyanoacetylene (\ce{DC3N}) has been detected in the ISM as well.
	The first astronomical observation of \ce{DC3N} has been reported towards TMC-1 \cite{langer1980tmc} by the detection of the $J=5\rightarrow4$ emission rotational transition around 42\,GHz.
	Consecutively, \ce{DC3N} has tentatively been detected in the high-mass star-forming regions Orion KL \cite{esplugues2013orion} and Sagittarius B2 \cite{belloche2016emoca}. In these regions, deuterium fractionation is not as effective as in dark clouds, thus preventing a strong enhancement above the deuterium cosmic abundance.
	Recently, \ce{DC3N} has been detected in some low-mass cores (see e.g., Refs.\cite{al2017iras},\cite{bianchi2019astrochemistry}) and in a sample of 15 high-mass star-forming cores \cite{rivilla2020dc3n}. The latter work, based on the spectroscopic results presented in this paper, suggests that \ce{DC3N} is enhanced in the cold and outer regions of star-forming regions, likely indicating the initial deuteration level of the large-scale molecular cloud within which star formation takes place. Rivilla \emph{et al.} \cite{rivilla2020dc3n} also summarize all the astronomical observations of \ce{DC3N} so far.
	
	Microwave (MW) transitions of \ce{DC3N} were first reported for the ground and the four lowest singly-excited states during the course of an extensive study of cyanoacetylene isotopologues \cite{mallinson1978microwave}.
	A larger number of vibrationally excited states was re-examined in depth some years later and a rigorous determination of the effective molecular parameters was attained \cite{plummer1988excited}.
	Recently, the laboratory investigation of the rotational spectrum of \ce{DC3N} has been extended to the THz regime for the ground and the $v_7=1$ states \cite{spahn2008thz}. In the same paper, the authors revised the \ce{^{14}N} and \ce{D} hyperfine-structure constants derived in Refs.~\cite{fliege1983hfs,tack1983beam} from supersonic-jet Fourier-Transform Microwave (FT-MW) spectroscopy.
	
	As far as its infrared (IR) spectrum is concerned, the experimental position and intensity of all fundamentals but the weak $\nu_4$ mode have been determined from low resolution (0.5\,\wn) studies \cite{uyemura1982ir,benilan2006ir}. Some combination and overtone bands were also observed in the same works.
	In addition, two medium resolution (0.025--0.050\,\wn) IR studies were performed by Mallinson \& Fayt \cite{mallinson1976high} and Coveliers \emph{et al.} \cite{coveliers1992far}. In the former, the band center of the three stretching modes of \ce{DC3N} ($\nu_1$, $\nu_2$, and $\nu_3$) has been determined; in the latter, the far-infrared (FIR) spectrum was recorded between 200 and 365\,\wn and the $\nu_7$ fundamental was analyzed together with the bands $\nu_6 - \nu_7$, $\nu_5 - \nu_7$, and $\nu_4 - \nu_6$, and their hot-bands.
	
	In this work, a detailed investigation of both millimeter/submillimeter-wave and infrared spectra of \ce{DC3N} is reported.
	Pure rotational transitions within all the vibrational states with energy lower than 1015\,\wn have been detected and 27 fundamental, overtone, combination, and hot ro-vibrational bands have been analyzed at high resolution (0.001--0.01\,\wn).
	The new measurements have been combined in a fit containing almost 6700 distinct transition frequencies, thus allowing the determination of a consistent set of spectroscopic parameters.
	This work represents the most exhaustive spectroscopic characterization of \ce{DC3N} so far and provides a robust line catalog useful for astronomical applications. Moreover, the large number of vibrational excited states are of interest for harmonic/anharmonic force field computations.
	
	The paper is structured as follows. First, the synthesis of the sample and the spectrometers used for spectral recording are described (\S\ref{sec:exp}). Then, the effective Hamiltonian employed for the energy levels description is given (\S\ref{sec:theory}). Successively, the general features of the spectra and their analysis are discussed (\S\ref{sec:analysis}).
	Finally, the results are summarized and the conclusions are presented (\S\ref{sec:concl}).

	\section{Experimental details}\label{sec:exp}
	
	\subsection{Synthesis of deuterocyanoacetylene}
	
	Methyl propiolate (\ce{HC+CCOOCH3}) was purchased from TCI-Europe and used without further purification.
	The \ce{DC3N} sample was synthesized in Rennes following the procedure described in Ref.~\cite{benilan2006ir}.
	Briefly, \ce{HC+CCOOCH3} was added dropwise to liquid ammonia resulting in a 100\,\% conversion into \ce{HC+CCONH2}.
	The propiolamide was then mixed with phosphorous anhydride (\ce{P4O10}) and calcined white sand; the whole system was heated up to 470\,K over 2\,h while connected to a liquid nitrogen-cooled trap where pure cyanoacetylene was collected.
	Cyanoacetylene (3\,g), heavy water (\ce{D2O}, 4\,mL) and potassium carbonate (\ce{K2CO3}, 50\,mg) were mixed together in an inert atmosphere. The biphasic mixture was then stirred for about 20\,min at room temperature.
	Subsequently, on a vacuum line, partially deuterated cyanoacetylene was condensed in a 77\,K cooled trap, while water was blocked in a first 220\,K trap. The operation was repeated 3 times by addition of \ce{D2O} and \ce{K2CO3} to the partially deuterated cyanoacetylene. The residual \ce{D2O} was removed by vaporisation on \ce{P4O10} and \ce{DC3N} was finally condensed in a trap cooled to 150\,K.
	Deuterocyanoacetylene with an isotopic purity greater than 98\,\% was obtained in a 67\,\% yield. The sample can be stored indefinitely at 250\,K without decomposition.
	
	\subsection{Infrared spectrometers}
	
	The FIR spectrum of \ce{DC3N} was recorded at the AILES beamline of the SOLEIL synchrotron facility using a Bruker IFS 125 FT interferometer \cite{Brubach2010} and a white-type multipass absorption cell whose optics were adjusted to obtain a 150\,m optical path length \cite{Pirali2012,Pirali2013}.
	For the present experiment, we used the far-IR synchrotron radiation continuum extracted by the AILES beamline.
	The interferometer was equipped with a 6\,\textmu m Mylar-composite beamsplitter and a 4\,K cooled \ce{Si}-bolometer.
	Two 50\,\textmu m-thick polypropylene windows isolated the cell from the interferometer, which was continuously evacuated to 0.01\,Pa limiting the absorption of atmospheric water. Vapor of \ce{DC3N} was injected into the absorption cell at a 25\,Pa pressure. The spectrum covers the range 70--500\,\wn and consists of the co-addition of 380 scans recorded at 0.00102\,\wn resolution.
	
	IR spectra in the 450--1600\,\wn range were recorded in Bologna using a Bomem DA3.002 Fourier-Transform spectrometer \cite{tamassia2020tfe}.
	It was equipped with a Globar source, a \ce{KBr} beamsplitter, and a liquid nitrogen-cooled \ce{HgCdTe} detector. A multi-pass cell with absorption-lengths from 4 to 8\,m was employed for the measurements.
	Sample pressures ranging between 25 and 650\,Pa were used to record the spectra.
	The resolution was generally 0.004\,\wn, except for the very weak $\nu_4$ band, which was recorded at a lower
	resolution of 0.012\,\wn.
	Several hundreds of scans, typically 800, were co-added in order to improve the signal-to-noise ratio (S/N) of the spectra.
	
	All the spectra have been calibrated using residual water or \ce{CO2} absorption lines whose reference wavenumbers were taken from Refs.~\cite{MatsushimaCalibrationH2O,Horneman2005} and from \texttt{HITRAN} \cite{gordon2017hitran2016}, respectively.
	No apodization functions were applied to the interferograms.
	
	\subsection{Millimeter and submillimeter spectrometers}
	
	Rotational spectra have been recorded using two frequency-modulation (FM) millimeter/submillimeter spectrometers located in Bologna and in Garching.
	
	The Bologna spectrometer has been described in details elsewhere \cite{degli2019determination,melosso2019sub}.
	Briefly, a Gunn diode oscillator operating in the W band (80--115\,GHz) was used as primary source
	of radiation, whose frequency and phase stability are ensured by a Phase-Lock Loop (PLL).
	Spectral coverage at higher frequencies was obtained by coupling the Gunn diode to passive frequency
	multipliers in cascade (doublers and triplers, Virginia Diodes, Inc.).
	The output radiation, sine-wave modulated in frequency ($f=48$\,kHz), was fed to the glass absorption cell
	containing \ce{DC3N} vapors at a pressure between 1 and 15\,Pa, depending on the
	intensity of the lines under consideration.
	The outcoming signal was detected by a Schottky barrier diode and sent to a Lock-in amplifier
	set at twice the modulation-frequency ($2f$ scheme); the demodulated signal is then filtered into a
	resistor-capacitor (RC) system before data acquisition.
	
	In Garching the CASAC spectrometer developed at the Max-Planck-Institut f\"ur extraterrestrische Physik was used.
	Full details on the experimental set-up are given in Ref.~\cite{bizzocchi2017accurate}; here, we report only a few key details
	which apply to the present investigation.
	The instrument is equipped with an active multiplier chain (Virginia Diodes) as a source of radiation in the 
	82--125\,GHz band.
	Further multiplier stages in cascade allow to extend the frequency coverage up to $\sim 1.1$\,THz with an 
	available power of 2--20\,\textmu W.
	The primary millimeter radiation stage is driven by a cm-wave synthesizer (Keysight E8257D) operating in the 
	18-28\,GHz band, which is locked to a \ce{Rb} atomic clock to achieve accurate frequency and phase stabilisation.
	A closed-cycle \ce{He}-cooled \ce{InSb} hot-electron bolometer operating at 4\,K (QMC) is used as a detector.
	As in Bologna, frequency ($f=50$\,kHz) modulation technique is employed and the second derivative of the actual
	absorption profile is thus recorded by the computer-controlled acquisition system after lock-in demodulation at $2f$\@.
	The absorption cell is a plain Pyrex tube (3\,m long and 5\,cm in diameter) fitted with high-density polyethylene
	windows.
	The measurement were performed using gaseous samples at pressure of a few Pa.
	In this condition, \ce{DC3N} is stable for ca.~2\,h without significant decomposition due to hydrogen exchange.
	
	The spectra were recorded in the frequency ranges 80--115\,GHz and 920--1070\,GHz in Garching, and in the window 240--440\,GHz in Bologna.
	
	\begin{table}[htb!]
		\centering
		\caption{Energy and intensity of all fundamental modes of \ce{DC3N}.}
		\label{tab:dc3n_modes}
		\scalebox{0.99}{
			\begin{threeparttable}
				\begin{tabular}{clccc}
					\hline\hline \\[-0.5ex]
					Modes & Description & Energy & Reference & \multicolumn{1}{c}{Abs. intensity}          \\[0.5ex]
					&             &    (\wn)     & & \multicolumn{1}{c}{(atm$^{-1}$cm$^{-2}$)} \\[0.5ex]
					\hline \\[-1.5ex]
					$\nu_1$ & \ce{C-D} stretching &  2608.520(3)   & \cite{mallinson1976high} & $81.3\pm5.7$\tnote{a}   \\[0.5ex]
					$\nu_2$ & \ce{C+C} stretching &  2252.155(3)   & \cite{mallinson1976high} & $50.5\pm2.4$\tnote{a}   \\[0.5ex]
					$\nu_3$ & \ce{C+N} stretching &  1968.329(3)   & \cite{mallinson1976high} & $38.7\pm4.0$\tnote{a}   \\[0.5ex]
					$\nu_4$ & \ce{C-C} stretching &  867.60(6)     & This Work                & $<0.1$\tnote{b}         \\[0.5ex]
					$\nu_5$ & \ce{CCD} bending   &  522.263933(7) & This Work                & $83.8\pm4.7$\tnote{a}   \\[0.5ex]
					$\nu_6$ & \ce{CCC} bending   &  492.759896(7) & This Work                & $106.\pm8$\tnote{a}     \\[0.5ex]
					$\nu_7$ & \ce{CCN} bending   &  211.550293(5) & This Work                & $0.89\pm0.11$\tnote{b}  \\[0.5ex]
					\hline\hline \\[-1ex]
				\end{tabular}
				\begin{tablenotes}
					\item \flushleft \textbf{[a]} From low-resolution integrated band-intensity measurements at 296\,K (Ref.~\cite{benilan2006ir}). \textbf{[b]} From low-resolution integrated band-intensity measurements at 293\,K (Ref.~\cite{uyemura1982ir}).
				\end{tablenotes}
			\end{threeparttable}
		}
	\end{table}

	\section{Theoretical background}\label{sec:theory}
	
	From a spectroscopic point of view, \ce{DC3N} is a closed-shell linear rotor.
	It has 7 vibrational modes: 4 stretchings ($\nu_1$--$\nu_4$; $\Sigma$ symmetry) and 3 doubly-degenerated bendings ($\nu_5$--$\nu_7$; $\Pi$ symmetry). They are summarized in Table~\ref{tab:dc3n_modes}.
	In the present work, only the low-lying vibrational states ($v_4$, $v_5$, $v_6$, and $v_7$, with one of multiple quanta of excitation) have been investigated for two main reasons:
	(i) transitions associated to the lower energy states are of astrophysical interest, and (ii) some of the vibrational states are connected by a network of anharmonic resonances fully described within our chosen energy threshold of 1015\,\wn; above this limit the states are either unperturbed or involved in higher-order resonances.
	Therefore, the stretching modes $\nu_1$, $\nu_2$, and $\nu_3$, lying above this threshold, have not been investigated.
	Conventionally, we labelled a given vibrational state with the notation $(v_4,v_5^{l_5},v_6^{l_6},v_7^{l_7})_{e/f}$, 
	where $l_t$ is the vibrational angular momentum quantum number associated to the bending mode $t$
	and the $e/f$ subscripts indicate the parity of the symmetrized wave functions \cite{brown1975labeling}.
	When the $l_t$ and $e/f$ labels are not indicated, we refer to all the possible sub-levels of a state.
	
	The full ro-vibrational wave-function is then given by the ket $\ket[e/f]{v_4,v_5^{l_5},v_6^{l_6},v_7^{l_7};J,k}$.
	The vibrational part of the wave-function is expressed as combination of one- or two-dimensional harmonic 
	oscillators, whereas the rotational part is the symmetric-top wave-function
	whose quantum number $k$ is given by
	$k = l_5 + l_6 + l_7$\@.
	A substate is denoted as $\Sigma$ when $k=0$, $\Pi$ for $\lvert k \rvert =1$, $\Delta$ for $\lvert k \rvert =2$, and so on.
	
	The following Wang-type linear combinations \cite{yamada1985effective} lead to symmetry-adapted basis functions:
	
	\begin{subequations} \label{eq:Wang}
		\begin{align}
			\left\lvert v_4,v_5^{l_5}\right.,&\left.v_6^{l_6},v_7^{l_7};J,k\right\rangle_{e/f}
			= \frac{1}{\sqrt{2}} \left\{
			\ket{v_4,v_5^{l_5},v_6^{l_6},v_7^{l_7};J,k} \pm (-1)^k
			\ket{v_4,v_5^{-l_5},v_6^{-l_6},v_7^{-l_7};J,-k} 
			\right\} \,,   \\
			\left\lvert v_4,0^0\right.,&\left.0^0,0^0;J,0\right\rangle_{e} = \ket{v_4,0^0,0^0,0^0;J,0} \,.
		\end{align}
	\end{subequations}
	
	The upper and lower signs ($\pm$) correspond to $e$ and $f$ wave-functions, respectively.
	For $\Sigma$ states ($k = 0$), the first non-zero $l_t$ is chosen positive.
	Here, the omission of the $e/f$ label indicates unsymmetrised wave-functions.
	The Hamiltonian used to reproduce the ro-vibrational energy levels is equivalent to the one used for \ce{HC3N} \cite{bizzocchi2017hc3n}:
	
	\begin{equation} \label{eq:hamdc3n}
		\mathscr{H} = \mathscr{H}_\mrm{rv} + \mathscr{H}_\mrm{\textit{l}-type} + \mathscr{H}_\mrm{res} \,,
	\end{equation}
	
	\noindent
	where $\mathscr{H}_\mrm{rv}$ is the ro-vibrational energy including centrifugal
	distortion corrections, $\mathscr{H}_\mrm{\textit{l}-type}$ represents the $l$-type interaction between the $l$
	sub-levels of the excited bending states, and $\mathscr{H}_\mrm{res}$ accounts for
	resonances among accidentally quasi-degenerate ro-vibrational states.
	The resonance network active in \ce{DC3N} resembles the one found for \ce{HC3N} and will be described later.
	
	The Hamiltonian matrix is built by using unsymmetrised ro-vibrational functions.
	It is subsequently factorized and symmetrized using Eqs.~\eqref{eq:Wang}.
	The matrix elements of the effective Hamiltonian are expressed using the formalism 
	already employed for the analysis of \ce{HC3N} \cite{bizzocchi2017hc3n}.
	
	\section{General features and analysis}\label{sec:analysis}
	
	\subsection{Vibrational spectra}
	
\begin{figure}[htb!]
		\centering
		\includegraphics[width=0.9\textwidth]{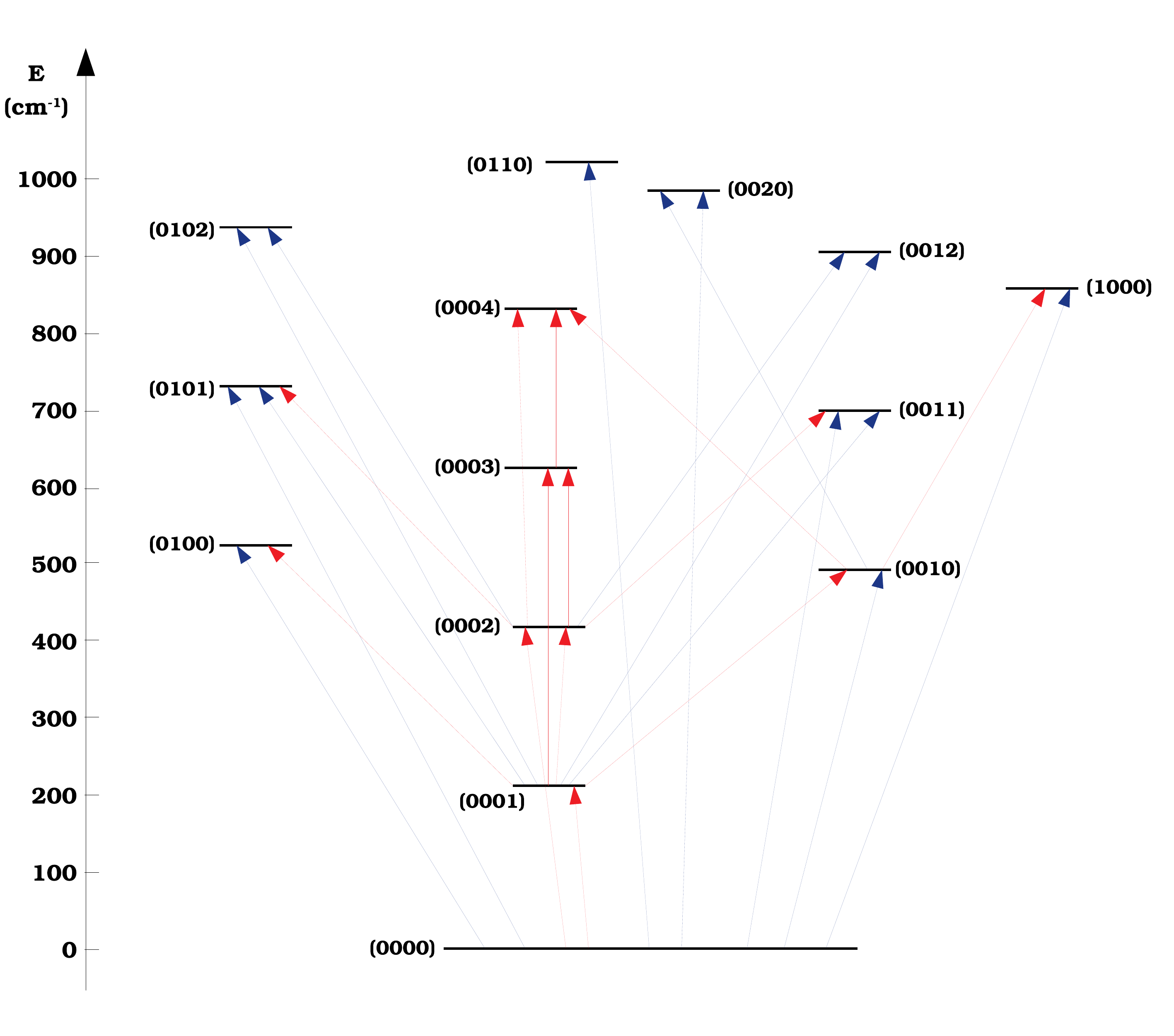}
		\caption{Vibrational energy-level diagram of \ce{DC3N} up to 1015\,\wn, where the arrows represent the 27 IR bands analyzed in this work. Red and blue arrows indicate the bands observed at SOLEIL and in Bologna, respectively.}
		\label{fig:dc3n_levels}
	\end{figure}

	Although infrared spectra were recorded up to 1600\,\wn in this study, our analysis is limited to the portion of the electromagnetic spectrum below $\sim$ 1040\,\wn. This is because the highest energy state within our threshold of 1015\,\wn is the (0110) state, whose combination band falls in the region 999--1035\,\wn.
	In total, 27 ro-vibrational bands have been observed at high resolution for the first time and successfully analyzed. They include fundamental, overtone, combination, and hot-bands, and are listed in Table~\ref{tab:dc3n_vibdata} along with the observed sub-bands, frequency and $J$ ranges, number of data used in the analysis, and the root-mean-square (\textit{rms}) error of the final fit.
	All the observed bands are also graphically displayed in Figure~\ref{fig:dc3n_levels}.
	
	\begin{table}[htb!]
		\centering
		\caption{Ro-vibrational bands recorded and analyzed in this work.}
		\label{tab:dc3n_vibdata}
		\begin{threeparttable}
			\begin{tabular}{llcccc}
				\hline\hline \\[-1ex]
				Band & Sub-bands & Freq. range & $J$ range & No. of lines & $rms \times 10^{4}$ \\[0.5ex]
				&           & (\wn)       &           &              & (\wn)                \\[0.5ex]
				\hline\ \\[-1.5ex]
				$\nu_7$                   & $\Pi - \Sigma^+$              & 190-240  & 1-93  & 258 & 0.5  \\[0.5ex]
				$\nu_6$                   & $\Pi - \Sigma^+$              & 466-522  & 2-109 & 267 & 3.6  \\[0.5ex]
				$\nu_5$                   & $\Pi - \Sigma^+$              & 500-557  & 0-117 & 255 & 3.7  \\[0.5ex]
				$\nu_4$                   & $\Sigma^+ - \Sigma^+$         & 830-865  & 0-61  & 109 & 9.6  \\[0.5ex]
				$2\nu_7$                  & $\Sigma^+ - \Sigma^+$         & 405-445  & 2-78  & 136 & 1.0  \\[0.5ex]
				$2\nu_6$                  & $\Sigma^+ - \Sigma^+$         & 975-1018 & 2-101 & 141 & 5.2  \\[0.5ex]
				$\nu_6+\nu_7$             & $\Sigma^+ - \Sigma^+$         & 686-736  & 1-89  & 166 & 2.5  \\[0.5ex]
				$\nu_5+\nu_7$             & $\Sigma^+ - \Sigma^+$         & 715-769  & 1-105 & 170 & 3.0  \\[0.5ex]
				$\nu_5+\nu_6$             & $\Sigma^+ - \Sigma^+$         & 999-1035 & 3-63  & 102 & 3.8  \\[0.5ex]
				$\nu_6 - \nu_7$           & $\Pi - \Pi$                   & 257-306  & 1-89  & 309 & 0.6  \\[0.5ex]
				$\nu_5 - \nu_7$           & $\Pi - \Pi$                   & 288-333  & 1-86  & 291 & 0.7  \\[0.5ex]
				$\nu_4 - \nu_6$           & $\Sigma^+ - \Pi$              & 329-375  & 1-79  & 222 & 0.6  \\[0.5ex]
				$4\nu_7 - \nu_6$          & $\Sigma^+ - \Pi$              & 351-352  & 59-60 &  2  & 0.4  \\[0.5ex]
				$2\nu_7 - \nu_7$          & $\Sigma - \Pi$                & 193-236  & 1-78  & 391 & 0.8  \\[0.5ex]
				$3\nu_7 - 2\nu_7$         & $\Pi - \Sigma^+$              & 191-235  & 4-77  & 178 & 0.7  \\[0.5ex]
				$4\nu_7 - 3\nu_7$         & $\Sigma^+ - \Pi$              & 193-219  & 10-58 & 89  & 0.9  \\[0.5ex]
				$3\nu_7 -  \nu_7$         & $\Pi - \Pi$                   & 405-441  & 4-68  & 208 & 1.0  \\[0.5ex]
				$\nu_6+\nu_7 - \nu_7$     & $(\Sigma, \Delta) - \Pi$      & 478-508  & 5-56  & 329 & 6.9  \\[0.5ex]
				$\nu_6+2\nu_7 - 2\nu_7$   & $\Pi - (\Sigma^{+}, \Delta)$  & 476-512  & 12-65 & 93  & 5.2  \\[0.5ex]
				$2\nu_6 - \nu_6$          & $\Sigma - \Pi$                & 500-519  & 10-68 & 43  & 7.0  \\[0.5ex]
				$\nu_5+\nu_7 - \nu_7$     & $(\Sigma^{+}, \Delta) - \Pi$  & 505-545  & 2-75  & 464 & 4.5  \\[0.5ex]
				$\nu_5+2\nu_7 - 2\nu_7$   & $\Pi - (\Sigma, \Delta)$      & 505-542  & 2-82  & 106 & 3.5  \\[0.5ex]
				$\nu_6+2\nu_7 - \nu_7$    & $\Pi - \Pi$                   & 690-718  & 4-53  & 153 & 4.7  \\[0.5ex]
				$\nu_5+2\nu_7 - \nu_7$    & $\Pi - \Pi$                   & 721-748  & 2-45  & 296 & 4.1  \\[0.5ex]
				$\nu_5+\nu_7 - 2\nu_7$    & $(\Sigma^{+}, \Delta) - (\Sigma^{+}, \Delta)$ & 290-330  & 2-81  & 261 & 0.9  \\[0.5ex]
				$\nu_6+\nu_7 - 2\nu_7$    & $(\Sigma^{+}, \Delta) - (\Sigma^{+}, \Delta)$ & 256-306  & 2-84  & 402 & 0.9  \\[0.5ex]
				$4\nu_7 - 2\nu_7$         & $(\Sigma^{+}, \Delta) - (\Sigma^{+}, \Delta)$ & 406-437  & 5-67  & 302 & 1.0  \\[0.5ex]
				\hline\hline
			\end{tabular}
		\end{threeparttable}
	\end{table}
	
	\begin{figure}[p!]
		\centering
		\includegraphics[width=0.95\textwidth]{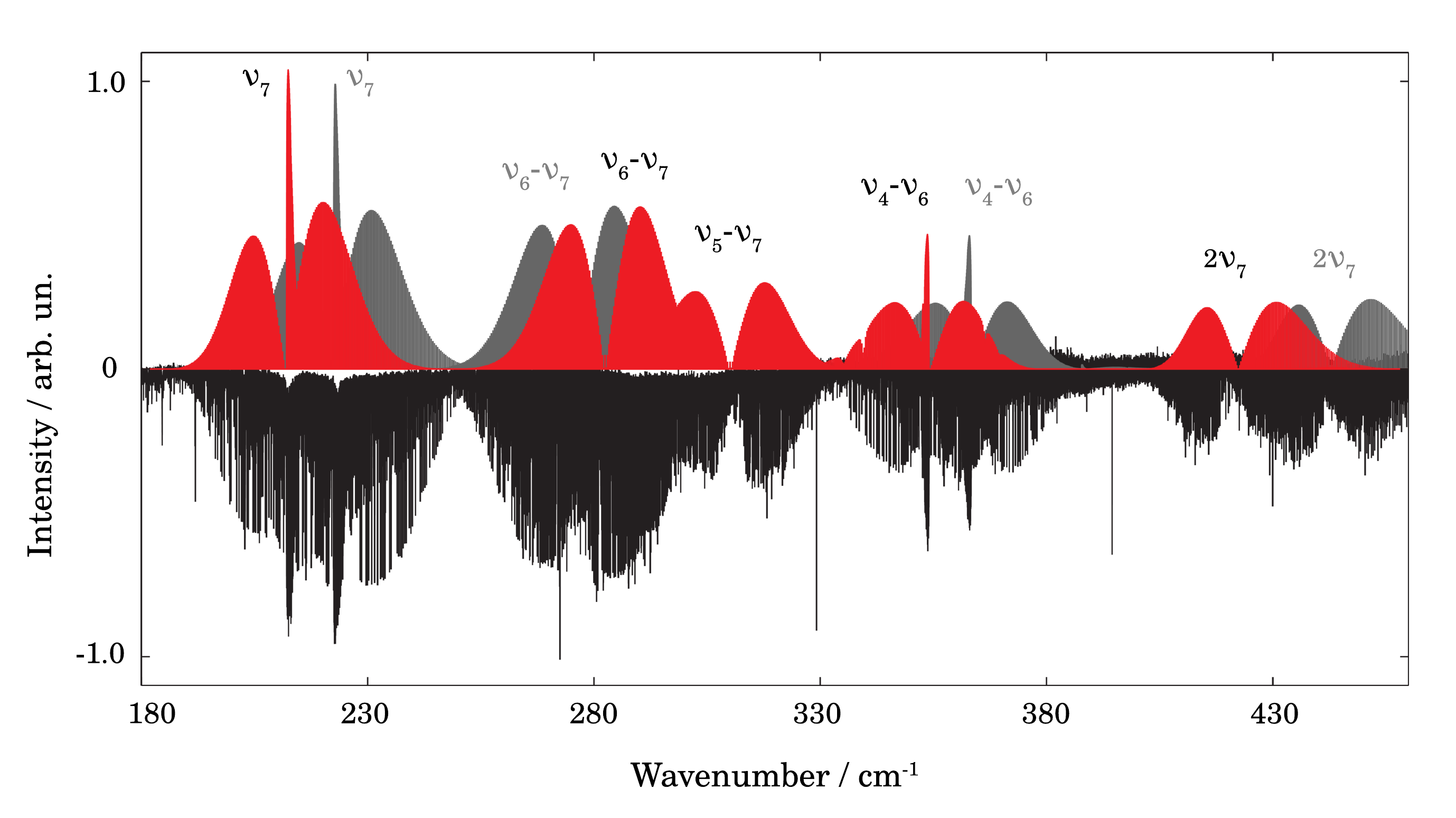}
		\includegraphics[width=0.95\textwidth]{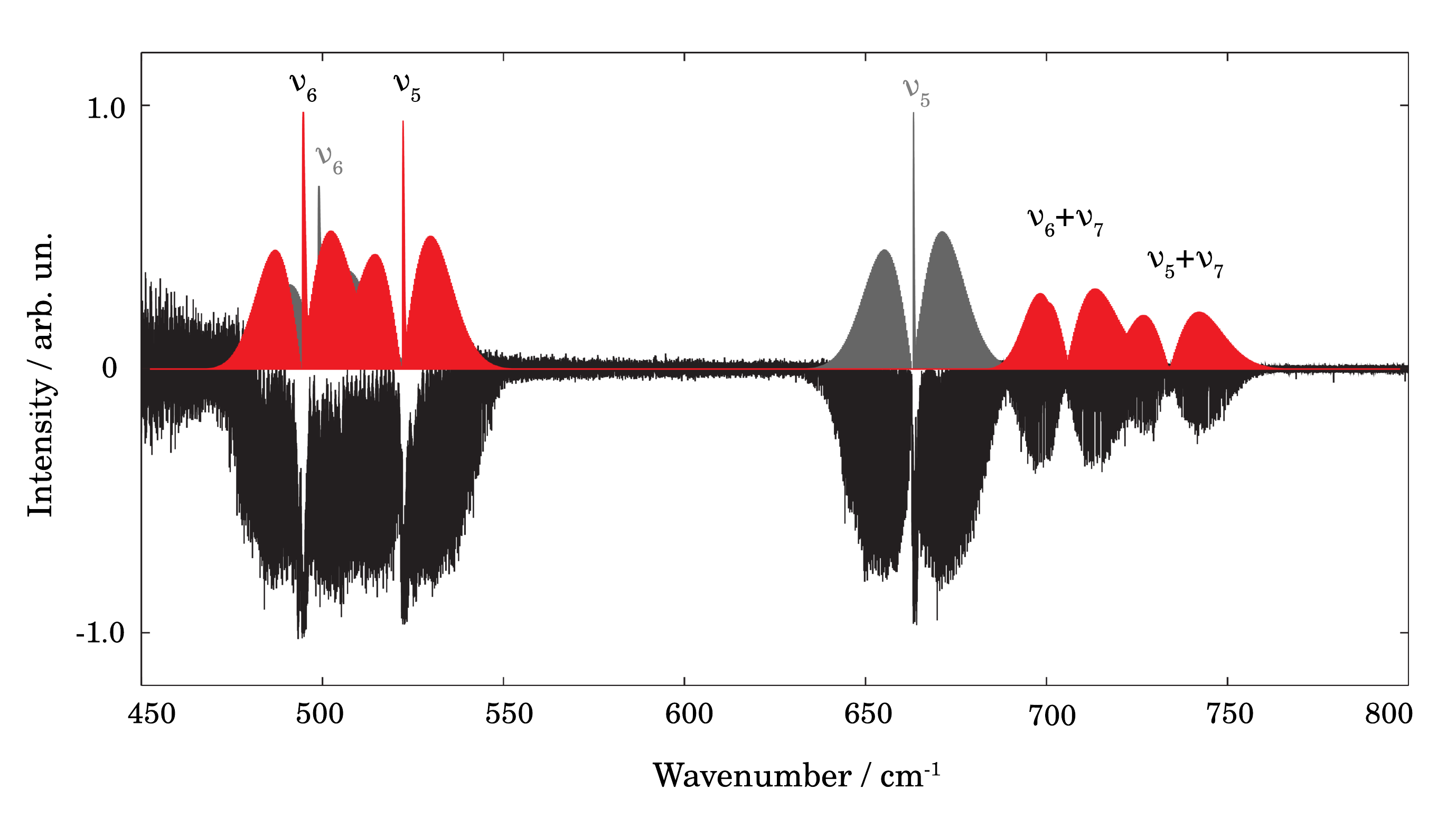}
		\caption{Portions of the FIR (upper panel) and MIR (lower panel) spectra of $d$-cyanoacetylene (black traces). A simulation spectrum of the most intense bands is also reported for both \ce{DC3N} (red) and \ce{HC3N} (grey). Lines belonging to \ce{CO} and various \ce{H2O} isotopologues were removed from the spectra.}
		\label{fig:dc3n_ir}
	\end{figure}
	
	Figure~\ref{fig:dc3n_ir} shows a general overview of portions of the FIR (180--460\,\wn range, upper panel) and mid-infrared (MIR, 450--800\,\wn range, bottom panel) spectra recorded in this work. 
	The most prominent bands in the FIR region are the $\nu_7$ fundamental, $\nu_6 - \nu_7$, $\nu_5 - \nu_7$, $\nu_4 - \nu_6$, and $2\nu_7$ overtone bands.
	The MIR region is dominated by the very strong fundamentals $\nu_6$ and $\nu_5$. The low-frequency side of the spectrum is particularly crowded due to the proximity of the two fundamentals, the presence of their associated hot-bands, and of the $\nu_6$ of \ce{HC3N} centered at 500\,\wn.
	\ce{HC3N} is present in the sample as result of the H/D exchange in the cell.
	
	Having a medium IR intensity, the combination bands $\nu_6+\nu_7$ and $\nu_5+\nu_7$ are well visible in the high-frequency part of the MIR spectrum as seen in the bottom panel of Figure~\ref{fig:dc3n_ir}.
	Although not displayed in Figure~\ref{fig:dc3n_ir}, the overtone $2\nu_6$ and the combination $\nu_5+\nu_6$ bands centered around 975--1018\,\wn and 999--1035\,\wn, respectively, are clearly detectable as well, despite the presence of strong absorption lines due to \ce{HDO}.
	The very weak ($< 0.1$ atm$^{-1}$ cm$^{-2}$) $\nu_4$ fundamental at 830--865\,\wn had to be recorded at higher pressure (400\,Pa) and lower resolution (0.012\,\wn). In this case, up to 2600 scans were co-added to improve the S/N of the spectrum.

	\subsection{Rotational spectra}
	
	Rotational spectra were recorded for all the 14 states whose vibrational energy do not exceed our threshold of 1015 \wn.
	Literature data were available for some of these states, as pointed out in Section~\ref{sec:intro}.
	However, line positions of some millimeter-wave transitions from Ref.~\cite{mallinson1978microwave} are affected by large uncertainties (up to 300\,kHz) and many data are limited to low frequencies.
	For these reasons, we decided to re-investigate and extend the spectrum for all these vibrational states.
	The largest improvements have been realized for the states (1000), (0110), (0020), and (0004) involved in a network of anharmonic resonances, for which extended data-sets were obtained.
	In particular, the (0110) state, not included in the analysis of Ref.~\cite{plummer1988excited},
	has been assigned for the first time in this study and its interaction with the (1000) state has been identified and properly accounted for.
	
	Table~\ref{tab:dc3n_rotdata} summarizes the set of rotational data used in the analysis, specifying the observed sub-levels, $J$ and frequency ranges, number of distinct fitted frequencies, the \textit{rms} error of the final fit, and the corresponding references used.
	
	\begin{table}[htb!]
		\centering
		\caption{Summary of the rotational data used in the analysis.}
		\label{tab:dc3n_rotdata}
		\begin{threeparttable}
			\begin{tabular}{lcccccl}
				\hline\hline \\[-1ex]
				State & $\left|k\right|$ & $J$ range & Freq. range & No. of lines & $rms$ & Reference \\[0.5ex]
				&     &           & (GHz)       &              & (kHz) &            \\[0.5ex]
				\hline\hline \\[-1.5ex]
				Ground state               & 0                       & 3-126 & 33-1069 & 52 & 13.2  & TW, Ma78, Pl88, Sp08  \\[0.5ex]
				$v_7 = 1$              & 1$_{\mrm{e,f}}$         & 5-105 & 50-896  & 67 & 10.9  & TW, Ma78, Pl88, Sp08  \\[0.5ex]
				$v_6 = 1$              & 1$_{\mrm{e,f}}$         & 7-44  & 67-381  & 42 & 17.0  & TW, Pl88              \\[0.5ex]
				$v_5 = 1$              & 1$_{\mrm{e,f}}$         & 7-44  & 67-381  & 42 & 13.7  & TW, Pl88              \\[0.5ex]
				$v_4 = 1$              & 0                       & 7-51  & 67-439  & 32 & 23.6  & TW, Pl88              \\[0.5ex]
				$v_7 = 2$              & 0, 2$_{\mrm{e,f}}$      & 7-44  & 67-383  & 61 & 24.0  & TW, Ma78, Pl88        \\[0.5ex]
				$v_7 = 3$              & (1, 3)$_{\mrm{e,f}}$    & 7-44  & 68-384  & 77 & 20.0  & TW, Ma78, Pl88        \\[0.5ex]
				$v_7 = 4$              & 0,(2, 4)$_{\mrm{e,f}}$  & 7-48  & 68-419  & 85 & 19.8  & TW, Pl88              \\[0.5ex]
				$v_6 = 2$              & 0, 2$_{\mrm{e,f}}$      & 7-44  & 67-382  & 54 & 44.7  & TW, Pl88              \\[0.5ex]
				$v_6 = v_7 = 1$    & (0, 2)$_{\mrm{e,f}}$    & 7-44  & 67-382  & 93 & 30.8  & TW, Ma78, Pl88        \\[0.5ex]
				$v_5 = v_7 = 1$    & (0, 2)$_{\mrm{e,f}}$    & 7-44  & 67-382  & 78 & 20.8  & TW, Pl88              \\[0.5ex]
				$v_5 = v_6 = 1$    & (0, 2)$_{\mrm{e,f}}$    & 9-44  & 84-381  & 63 & 14.6  & TW                    \\[0.5ex]
				$v_6 = 1, v_7 = 2$ & $(\pm1, 3)_{\mrm{e,f}}$ & 7-46  & 68-400  & 95 & 21.9  & TW, Pl88              \\[0.5ex]
				$v_5 = 1, v_7 = 2$ & $(\pm1, 3)_{\mrm{e,f}}$ & 9-44  & 85-383  & 97 & 17.4  & TW                    \\[0.5ex]
				interstate\tnote{a}        &                         & 44-49 & 364-429 & 10 & 19.7  & TW                    \\[0.5ex]
				\hline\hline
			\end{tabular}
			\begin{tablenotes}
				\item \flushleft Abbreviations are used as follow: \textbf{TW} This work, \textbf{Ma78} Mallinson \& De Zafra (1978) \cite{mallinson1978microwave}, \textbf{Pl88} Plummer \emph{et al.} (1988) \cite{plummer1988excited}, \textbf{Sp08} Spahn \emph{et al.} (2008) \cite{spahn2008thz}. [a] Transitions between the interacting states (1000) and (0004).
			\end{tablenotes}
		\end{threeparttable}
	\end{table}
	
\begin{figure}[htb!]
		\centering
		\includegraphics[width=\textwidth]{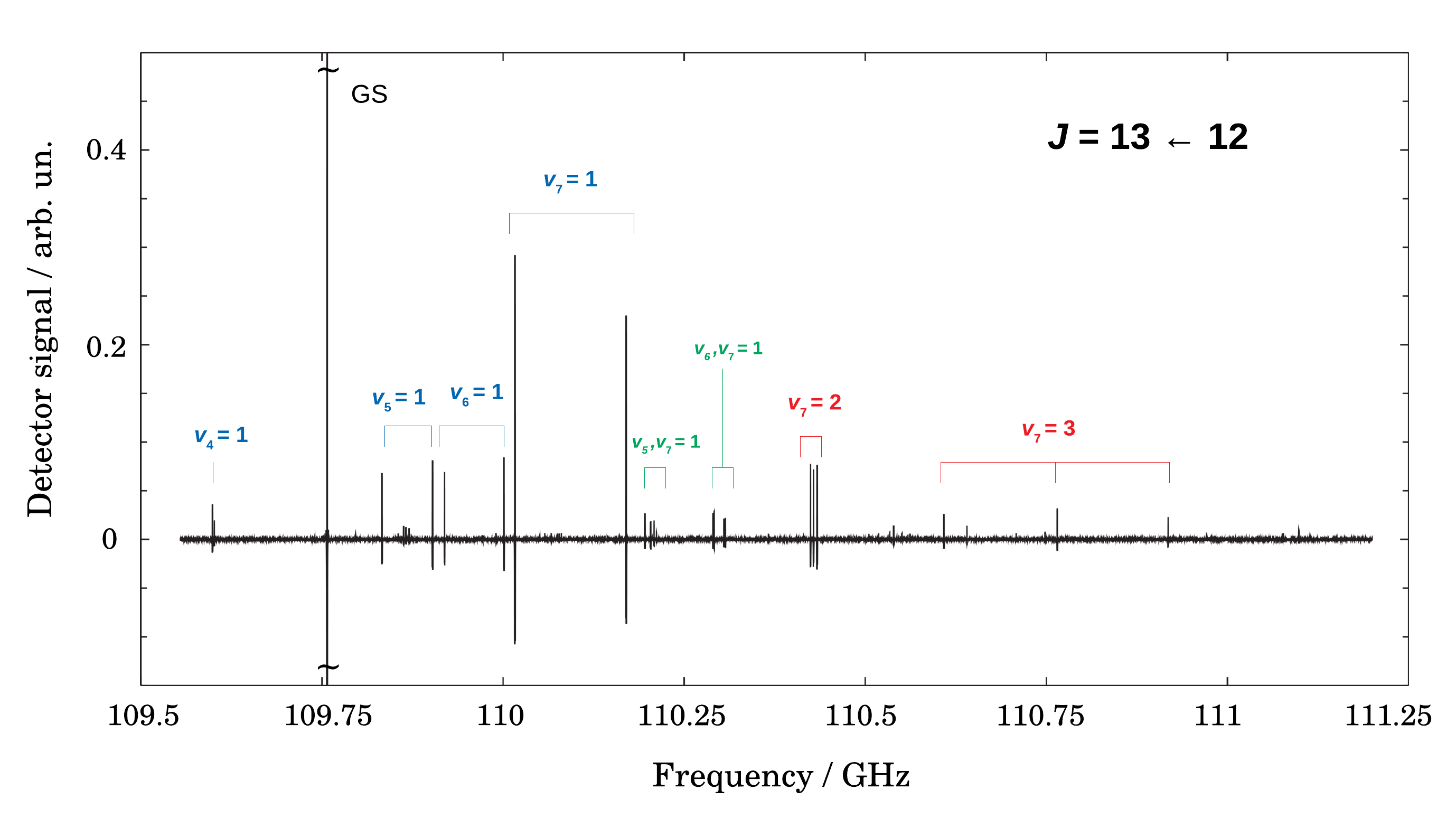}
		\caption{A 2\,GHz broad scan of the $J = 13\leftarrow 12$ rotational transition of \ce{DC3N} around 110\,GHz. The spectrum was recorded at room temperature, with \ce{DC3N} at a pressure of 0.05\,Pa, RC= 3\,ms, frequency step 50\,kHz, FM = 120\,kHz, scan speed = 0.4\,MHz/s, 2 scans. The arbitrary units of the $y$-axis are set so that the intensity of the ground state (GS) transition, out of scale in the figure, is 1.}
		\label{fig:dc3n_long}
	\end{figure}

	With the exception of $v_4 = 1$, all the states possess a rotational constant $B$ greater than that of the ground state and therefore their rotational lines lie at frequencies higher than those of the corresponding ground state transition.
	This can be seen in Figure~\ref{fig:dc3n_long}, where the broad scan covers the $J=13 \leftarrow 12$ transitions for many vibrational satellites.
	In this excerpt, the $l$-type resonance patterns of all the excited bending states analyzed are visible.
	From a visual inspection, it is easy to associate some of these patterns to the pertaining state: the ground and $v_4=1$ exhibit a single line, while each bending state has $\sum=\prod_{t}(l_t+1)$ lines (even though not always resolvable).
	
	\subsection{Analysis of the spectra}
	
	The sample of pure rotational and ro-vibrational data contains 6691 distinct frequencies involving 14 vibrational states of \ce{DC3N}. This work represents the first-ever investigation of its ro-vibrational spectrum in the region between 365 and 1040\,\wn.
	Moreover, the FIR spectrum has been thoroughly re-investigated at higher resolution with an accuracy two or three orders of magnitude better than Ref.~\cite{coveliers1992far}.
	As far as the rotational spectrum is concerned, this work extends the observation of excited states transitions to the submillimeter-wave region.
	In addition, rotational transitions with $J$ up to 126 were recorded at THz frequencies (1.069\,THz) although only for the ground state.
	
	In the combined fit, a different weight was given to each datum in order to take into account the different measurements precision.
	Uncertainties spanning from 0.0004 to 0.00075\,\wn were used for the infrared measurements performed in Bologna; the weak $\nu_4$ band being the only exception, for which an uncertainty of 0.001\,\wn was assumed. FIR data recorded at higher resolution with the FT-IR spectrometer of the AILES beamline have been given uncertainties between 0.00005 and 0.0001\,\wn, based on calibration residuals and the S/N of spectral lines.
	As far as pure rotational transitions are concerned, we assumed a typical experimental error of 10--20\,kHz for our new millimeter/submillimeter measurements. Data from literature were used with the uncertainty stated in the original papers \cite{mallinson1978microwave,plummer1988excited,spahn2008thz}.
	Only few lines from Ref.~\cite{mallinson1978microwave}, whose residuals were far off their declared errors, were not used in the fit.
	
	\begin{table}[htb!]
		\centering
		\caption{Spectroscopic constants derived for \ce{DC3N} in the ground and $v_4 = 1$ states.}
		\label{tab:dc3n_gs}
		\begin{threeparttable}
			\begin{tabular}{ll..}
				\hline\hline \\[-1ex]
				Constant & Unit & \multicolumn{1}{c}{Ground state} & \multicolumn{1}{c}{$v_4 = 1$} \\[0.5ex]
				\hline \\[-1.5ex]
				$G_v$  &  \wn  &    0.0           &  867.594(75)        \\[0.5ex]
				$B_v$  &  MHz  & 4221.580853(37)  & 4212.271(16)        \\[0.5ex]
				$D_v$  &  kHz  &    0.4517857(89) &    0.45312(11)     \\[0.5ex]
				$H_v$  &  mHz  &    0.03949(78)   &    0.03949\tnote{a} \\[0.5ex]
				$L_v$  &  nHz  &   -0.154(23)     &   -0.154\tnote{a}   \\[0.5ex]
				\hline\hline
			\end{tabular}
			\begin{tablenotes}
				\item \flushleft Number in parenthesis are one standard deviation in units of the last quoted digit. [a] Kept fixed to ground state value.
			\end{tablenotes}
		\end{threeparttable}
	\end{table}
	
	
	\begin{table}[htb!]
		\centering
		\caption{Spectroscopic constants derived for \ce{DC3N} in singly-excited bending states.}
		\label{tab:dc3n_singly}
		\begin{threeparttable}
			\begin{tabular}{ll...}
				\hline\hline \\[-1ex]
				Constant & Unit & \multicolumn{1}{c}{$v_7 = 1$} & \multicolumn{1}{c}{$v_6 = 1$} & \multicolumn{1}{c}{$v_5 = 1$} \\[0.5ex]
				\hline \\[-1.5ex]
				$G_v$                       & \wn &  211.5502859(33)  &  492.7605681(48)  &  522.2639331(49)     \\[0.5ex]
				$X_{\mrm{L}_\mrm{(tt)}}$ & GHz &  19.5125\tnote{a} &   56.39\tnote{a}  & ...                  \\[0.5ex]
				$B_v$                       & MHz & 4234.519466(31)   & 4229.25208(11)    &  4225.835835(71)     \\[0.5ex]
				$D_v$                       & kHz &    0.4718865(60)  &    0.462123(45)   &     0.452490(14)     \\[0.5ex]
				$H_v$                       & mHz &    0.08240(30)    &    0.0637(45)     &     0.03949\tnote{b} \\[0.5ex]
				$L_v$                       & nHz &   -0.154\tnote{b} &   -0.154\tnote{b} &    -0.154\tnote{b}   \\[0.5ex]
				$d_{\mrm{JL}_\mrm{(tt)}}$   & kHz &   -9.971\tnote{a} &    141.5\tnote{a} &  ...                 \\[0.5ex]
				$q_{\mrm{t}}$               & MHz &    5.907823(56)   &    3.15095(11)    &  2.68903(13)         \\[0.5ex]
				$q_{\mrm{tJ}}$              & Hz  &  -13.646(11)      &   -1.572(22)      & -1.627(27)           \\[0.5ex]
				$q_{\mrm{tJJ}}$             & \textmu Hz &   43.37(57)       &  ...              & ...                  \\[0.5ex]
				\hline\hline
			\end{tabular}
			\begin{tablenotes}
				\item \flushleft Number in parenthesis are one standard deviation in units of the last quoted digit. [a] Constrained value, see text. [b] Kept fixed to ground state value.
			\end{tablenotes}
		\end{threeparttable}
	\end{table}
	
	
	\begin{table}[htb!]
		\centering
		\caption{Spectroscopic constants derived for \ce{DC3N} in overtone states.}
		\label{tab:dc3n_over}
		\begin{threeparttable}
			\begin{tabular}{llmmmm}
				\hline\hline \\[-1ex]
				Constant & Unit & \multicolumn{1}{c}{$v_7 = 2$} & \multicolumn{1}{c}{$v_7 = 3$} & \multicolumn{1}{c}{$v_7 = 4$} & \multicolumn{1}{c}{$v_6 = 2$} \\[0.5ex]
				\hline \\[-1.5ex]
				$G_v$                       & \wn &  422.3753581(61)    &  632.510162(64)     &   841.9860892(95)    &  983.021(78)        \\[0.5ex]
				$X_{\mrm{L}_\mrm{(tt)}}$ & GHz &   19.354043(62)     &   19.1988(19)       &    19.03874(52)     &   56.39(58)         \\[0.5ex]
				$y_{\mrm{L}_\mrm{(tt)}}$    & MHz &   ...               &   ...               &     1.82(12)        &   ...               \\[0.5ex]
				$B_v$                       & MHz & 4247.45224(11)      & 4260.38118(15)      & 4273.30576(21)      & 4236.558(16)        \\[0.5ex]
				$D_v$                       & kHz &    0.491827(21)     &    0.512469(34)     &    0.53448(11)      &    0.471993(61)     \\[0.5ex]
				$H_v$                       & mHz &    0.03949\tnote{b} &    0.03949\tnote{b} &    0.176(21)        &    0.03949\tnote{b} \\[0.5ex]
				$L_v$                       & nHz &   -0.154\tnote{b}   &   -0.154\tnote{b}   &   -0.154\tnote{b}   &   -0.154\tnote{b}   \\[0.5ex]
				$d_{\mrm{JL}_\mrm{(tt)}}$   & kHz &  -10.426(30)        &  -10.947(24)        &   -11.368(21)       &  141.6(39)          \\[0.5ex]
				$h_{\mrm{JL}_\mrm{(tt)}}$   &  Hz &   ...               &   ...               &     -0.0552(64)     &   ...               \\[0.5ex]
				$q_{\mrm{t}}$               & MHz &    5.93258(10)      &    5.95888(12)      &    5.98281(15)      &    3.15095\tnote{a} \\[0.5ex]
				$q_{\mrm{tJ}}$              & Hz  &  -13.897\tnote{a}   &  -14.149(36)        &  -14.288(46)        &   -1.571\tnote{a}   \\[0.5ex]
				$q_{\mrm{tJJ}}$             & \textmu Hz &   43.37\tnote{a}    &   43.37\tnote{a}    &   43.37\tnote{a}    & ...                 \\[0.5ex]
				\hline\hline
			\end{tabular}
			\begin{tablenotes}
				\item \flushleft Number in parenthesis are one standard deviation in units of the last quoted digit. [a] Constrained value, see text. [b] Kept fixed to ground state value.
			\end{tablenotes}
		\end{threeparttable}
	\end{table}
	
	\begin{sidewaystable}[ph!]
		\centering
		\caption{Spectroscopic constants derived for \ce{DC3N} in combination states.}
		\label{tab:dc3n_comb}
		\begin{threeparttable}
			\begin{tabular}{llmmmmm}
				\hline\hline \\[-1ex]
				Constant & Unit & \multicolumn{1}{c}{$v_6 = v_7 = 1$} & \multicolumn{1}{c}{$v_5 = v_7 = 1$} & \multicolumn{1}{c}{$v_5 = v_6 = 1$} & \multicolumn{1}{c}{$v_6 = 1, v_7 = 2$} & \multicolumn{1}{c}{$v_5 = 1, v_7 = 2$} \\[0.5ex]
				\hline \\[-1.5ex]
				$G_v$                       & \wn &  703.8550157(95)   &  734.058721(13)  &  1014.2947(11)     &  914.21137(23)      &  945.143633(32)     \\[0.5ex]
				$X_{\mrm{L}_\mrm{(aa)}}$ & GHz &   56.39\tnote{a}  &   ...            &   ...               &   56.39\tnote{a}    &   ...               \\[0.5ex]
				$X_{\mrm{L}_\mrm{(bb)}}$ & GHz &  19.3189\tnote{a} & 19.5125\tnote{a} &     56.39\tnote{a}  &   19.1254(87)       &   19.3142(16)       \\[0.5ex]
				$X_{\mrm{L}_\mrm{(ab)}}$ & GHz &   16.16651(21)    &   23.13131(38)   &     40.245(33)      &   16.2848(57)       &   23.0216(30)       \\[0.5ex]
				$r_{\mrm{ab}}$            & GHz &  -17.04625(41)    &    0.32219(69)   &    -63.498(67)      &  -16.6526(81)       &    0.87043(34)      \\[0.5ex]
				$r_{\mrm{abJ}}$           & kHz &   -5.784(70)      &  -65.965(72)     &   ...               &  -12.0(12)          &  -64.05(11)        \\[0.5ex]
				$B_v$                       & MHz & 4242.274182(83)   & 4238.741823(94)  & 4233.55575(16)      & 4255.2987(20)       & 4251.64459(18)      \\[0.5ex]
				$D_v$                       & kHz &    0.481887(24)   &    0.472594(28)  &    0.463447(40)     &    0.502237(37)     &    0.492993(33)     \\[0.5ex]
				$H_v$                       & mHz &  0.03949\tnote{b} & 0.03949\tnote{b} &    0.03949\tnote{b} &    0.03949\tnote{b} &    0.03949\tnote{b} \\[0.5ex]
				$L_v$                       & nHz &   -0.154\tnote{b} &  -0.154\tnote{b} &   -0.154\tnote{b}   &   -0.154\tnote{b}   &   -0.154\tnote{b}   \\[0.5ex]
				$d_{\mrm{JL}_\mrm{(aa)}}$   & kHz &  -11.254\tnote{a} &  ...             &   ...               &  141.5\tnote{a}     &  ...                \\[0.5ex]
				$d_{\mrm{JL}_\mrm{(bb)}}$   & kHz &  141.5\tnote{a}   &  -9.971\tnote{a} &  141.5\tnote{a}     &  -12.54(76)         &  -10.482(55)        \\[0.5ex]
				$d_{\mrm{JL}_\mrm{(ab)}}$   & kHz &   43.88(12)       &   -5.08(13)      &   80.35(29)         &   43.75(46)         &   -6.810(46)        \\[0.5ex]
				$q_{\mrm{a}}$               & MHz &    3.17827(15)    &    2.70634(31)   &    2.69091(31)      &    3.19092(20)      &    2.72654(16)      \\[0.5ex]
				$q_{\mrm{aJ}}$              & Hz  &   -1.571\tnote{a} &  -1.626\tnote{a} &   -1.626\tnote{a}   &   -1.517\tnote{a}   &   -1.626\tnote{a}   \\[0.5ex]
				$q_{\mrm{b}}$               & MHz &    5.94427(18)    &    5.90943(73)   &    3.15095\tnote{a} &    5.9489(16)       &    5.93169(23)      \\[0.5ex]
				$q_{\mrm{bJ}}$              & Hz  &  -13.646\tnote{a} &  -13.738(98)      &   -1.571\tnote{a}   &  -14.17(33)         &  -13.646\tnote{a}   \\[0.5ex]
				$q_{\mrm{bJJ}}$             & \textmu Hz &   43.37\tnote{a}  &   43.37\tnote{a} &   ...               &   43.37\tnote{a}    &   43.37\tnote{a}    \\[0.5ex]
				$u_{\mrm{ab}}$              & Hz  &   ...             &   ...            &   -1.641(74)        & ...                 & ...                 \\[0.5ex]
				\hline\hline
			\end{tabular}
			\begin{tablenotes}
				\item \flushleft Number in parenthesis are one standard deviation in units of the last quoted digit. [a] Constrained value, see text. [b] Kept fixed to ground state value.
			\end{tablenotes}
		\end{threeparttable}
	\end{sidewaystable}

	The spectral analysis was performed using a custom \texttt{PYTHON} code that employs the \texttt{SPFIT} program \cite{pickett1991} as computational core (see Ref.~\cite{bizzocchi2017hc3n} for further details about the code).
	The data were fitted to the Hamiltonian of Eq.~\eqref{eq:hamdc3n} and its coefficients optimized in an iterative least-squares procedure.
	Some spectroscopic parameters could not be determined from the available experimental data.
	In these cases, the constant of a given vibrational level were derived from the corresponding optimized values obtained for other levels belonging to the same vibrational manifold considering, whenever feasible, a vibrational dependence.
	In other cases, they were simply fixed to zero.
	The spectroscopic parameters obtained from the combined fit procedure are collected in Tables~\ref{tab:dc3n_gs}-\ref{tab:dc3n_comb}.
	
	As anticipated, the analysis of \ce{DC3N} follows the approach successfully adopted for \ce{HC3N} \cite{bizzocchi2017hc3n}.
	The main difference is the set-up of the anharmonic resonances network, which arises from the different energy of some vibrational levels due to the isotopic substitution. In particular, the $\nu_5$ vibrational energy, 663.36848(3)\,\wn in \ce{HC3N}, drops to 522.26378(2)\,\wn in \ce{DC3N}. For \ce{HC3N}, two resonant systems were described: (i) $v_5=1 \sim v_7=3$ and ii) $v_4=1 \sim v_5=v_7=1 \sim v_6=2 \sim v_7=4$. Of the two systems, the former is not present in \ce{DC3N} while the latter is almost the same, except for $v_5=v_7=1$, replaced by $v_5=v_6=1$. The treatment of such perturbations led to the determination of the corresponding interaction parameters, $C_{30}$ for the cubic terms, ($v_4=1$)--($v_6=2$) and ($v_4=1$)--($v_5=v_6=1$), and $C_{50}$ for the quintic term ($v_4=1$)--($v_7=4$). Moreover, a centrifugal distortion parameter $C_{50}^J$ was included in the analysis.
	
	For the states involved in this resonance system, many experimental data are available.
	In the MIR region, we recorded the $\nu_4$, $\nu_5+\nu_6$, and $2\nu_6$ bands that provide the energy position
	for most of the interacting levels.
	The energy of the $v_7=4$ was determined through the FIR spectrum, where the $4\nu_7 \leftarrow 3\nu_7$ hot-band and the $4\nu_7 \leftarrow \nu_6$ band were detected.
	A large pure rotational data-set is also available for the polyad of interacting states.
	Besides several rotational transitions observed within the vibrational states, a small set of interstate transitions between the (1000) and (0004) states were identified.
	The coefficients $C_{mn}$ of the resonance Hamiltonian are given in Table~\ref{tab:dc3n_reso}.
	
	\begin{table}[htb!]
		\centering
		\caption{Resonance parameters.}
		\label{tab:dc3n_reso}
		\begin{threeparttable}
			\begin{tabular}{lcl.}
				\hline\hline \\[-1ex]
				Interacting states & Parameter & Unit & \multicolumn{1}{c}{Value} \\[0.5ex]
				\hline \\[-1.5ex]
				($v_4 = 1$) -- ($v_6 = 2$)           & $C_{\mrm{30}}$   & \wn & 17.422(33)   \\[0.5ex]
				($v_4 = 1$) -- ($v_5 = v_6 = 1$) & $C_{\mrm{30}}$   & \wn & -6.527(13)   \\[0.5ex]
				($v_4 = 1$) -- ($v_7 = 4$)           & $C_{\mrm{50}}$   & GHz &  2.70065(76) \\[0.5ex]
				& $C_{\mrm{50}}^J$ & kHz & 9.807(35)    \\[0.5ex]
				\hline\hline
			\end{tabular}
			\begin{tablenotes}
				\item \flushleft Number in parenthesis are one standard deviation in units of the last quoted digit.
			\end{tablenotes}
		\end{threeparttable}
	\end{table}

	
	\section{Conclusions}\label{sec:concl}
	
	In this work, a large set of high-resolution rotational and ro-vibrational data of \ce{DC3N} has been recorded and analyzed in order to achieve a detailed knowledge of all the vibrational states approximately below 1000\,\wn of energy.
	To reach this goal, infrared spectra of \ce{DC3N} have been recorded in the range 150--1600\,\wn at high resolution (0.001--0.004\,\wn). In this region, 27 fundamental, overtone, combination, and hot-bands have been observed and analyzed.
	Notably, the very weak $\nu_4$ fundamental has also been detected, even though at lower resolution (0.012\,\wn).
	Also, pure rotational transitions for 14 states have been recorded to extend the investigation of the spectrum to the submillimeter-wave region up to \emph{ca.} 500\,GHz.
	
	Almost 6700 experimental transitions were included in a least-squares fit procedure thanks to which a large number of rotational and ro-vibrational spectroscopic parameters have been determined for 14 different vibrational states. The whole set of data has been fitted with an overall weighted standard deviation $\sigma$ of 0.95, meaning that on average all data are well-reproduced within their given uncertainties.
	The vibrational energies were determined experimentally for all the investigated states, without any assumption.
	The combination of both high-resolution ro-vibrational data and pure rotational measurements allowed an accurate modeling of the spectrum of \ce{DC3N}, including perturbations produced by the observed anharmonic resonances.
	The interaction between the (1000) and (0110) states has been introduced for the first time, with the effect to eliminate the residual discrepancies described in Refs.~\cite{plummer1988excited,coveliers1992far}.
	
	The present work shows once again the success of a combined analysis of data from different spectral regions, like infrared and millimeter-wave fields. The results are generally more coherent and fewer assumptions are needed, if not any. Also, a more extended set of spectroscopic parameters can be obtained with reliability.
	
	This study provides an extensive line catalog (deposited as Supplementary Material) which can be used to assist future astronomical observations of \ce{DC3N} and is suitable for modeling both cold and hot regions of the interstellar medium.

	\section{Acknowledgement}
	
	This study was supported by Bologna University (RFO funds) and by MIUR (Project PRIN 2015: STARS in the CAOS, Grant Number 2015F59J3R). This work has been performed under the SOLEIL proposal \#20190128; we acknowledge the SOLEIL facility for provision of synchrotron radiation and would like to thank the AILES beamline staff for their assistance.
	L.B., P.C., and B.M.G. acknowledge the support by the Max Planck Society.
	V.M.R. has received funding from the European Union's Horizon 2020 research and innovation programme under the Marie Sk\l{}odowska-Curie grant agreement No 664931. LC acknowledges support from the Italian Ministero dell’Istruzione, Universit\`a e Ricerca through the grant Progetti Premiali 2012 - iALMA (CUP C52I13000140001).
	J.-C.G. thanks the Centre National d'Etudes Spatiales (CNES) for a grant.
	
	\bibliography{bibfile}
	
\end{document}